\documentclass[aps, pra, a4paper, amsfonts, amssymb, amsmath, reprint, showkeys,superscriptaddress, nofootinbib]{revtex4-2}

\usepackage[normalem]{ulem}
\usepackage[colorlinks=true,allcolors=blue,bookmarks=false,pdfusetitle]{hyperref}
\usepackage{url}
\usepackage{cleveref} 
\usepackage{mathtools}
\usepackage{graphicx}
\usepackage{array}[=2016-10-06]

\usepackage{amsthm}
\usepackage{physics}
\usepackage{xcolor}
\usepackage{adjustbox}
\usepackage{placeins}
\usepackage[T1]{fontenc}
\usepackage{lipsum}
\usepackage{csquotes}

\usepackage[left=23mm,right=13mm,top=35mm,columnsep=15pt]{geometry} 

\usepackage[colorlinks=true,allcolors=blue,bookmarks=false,pdfusetitle]{hyperref} 
\newcolumntype{C}[1]{>{\centering\arraybackslash}m{#1}}

\begin{document}

\title{Bell inequality violation with momentum-entangled massive particles}

\author{Y. S. Athreya} 
\affiliation{%
	Research School of Physics, Australian National University, Canberra 2601, Australia
}

\author{S. Kannan}
\affiliation{%
	Research School of Physics, Australian National University, Canberra 2601, Australia
}
\author{X. T. Yan}
\affiliation{%
	Research School of Physics, Australian National University, Canberra 2601, Australia
}

\author{K. V. Kheruntsyan}
\affiliation{%
	School of Mathematics and Physics, University of Queensland, Brisbane, Queensland 4072, Australia
}
\author{A. G. Truscott}
\affiliation{%
	Research School of Physics, Australian National University, Canberra 2601, Australia
}
\author{S. S. Hodgman}
\email{sean.hodgman@anu.edu.au}
\affiliation{%
	Research School of Physics, Australian National University, Canberra 2601, Australia
}

\date{\today} 

\begin{abstract}

Bell's theorem revealed the fundamental incompatibility between the predictions of quantum mechanics and local realism. Bell inequality violations have since demonstrated quantum nonlocality using photons and the internal states of massive particles, but never using their motional states. Here we report the first Bell inequality violation in the motional states of massive particles. Using momentum-entangled pairs of metastable helium atoms manipulated by matter-wave interferometry, we measure a Clauser–Horne–Shimony–Holt (CHSH) Bell parameter of $S=2.52\pm0.17$, violating the CHSH-Bell inequality ($S\le2$). Our work completes a long-standing objective in quantum atom optics by extending Bell tests from internal quantum variables to the external degrees of freedom of massive particles, opening a new regime for exploring quantum nonlocality in matter waves and for investigating the interplay between quantum mechanics and gravity.

\end{abstract}


\maketitle


Bell’s theorem established that no theory based on local realism can reproduce all predictions of quantum mechanics, transforming the debate over the completeness of quantum theory into an experimentally testable question \cite{Bell1964,Bell:book,Clauser1969,aspect2004bellstheoremnaive}. Over the past five decades, Bell inequality violations have become one of the defining demonstrations of quantum nonlocality, first with entangled photons \cite{Freedman1972,aspect1982ExperimentalTestBells,LoopholeFreeBellPhotons2015,LoopholeFreeLocalRealism2015}  and later with a wide range of massive quantum systems, including trapped ions \cite{rowe2001experimental}, neutral atoms ~\cite{hofmann2012heralded}, superconducting circuits \cite{ansmann2009violation}, and solid-state spin qubits \cite{hensen2015loophole}. 
However, previous Bell inequality violations involving massive particles have encoded the measurement observables in spin or \emph{internal} degrees of freedom.
The translational motion of massive particles---the defining \emph{external} degree of freedom of matter waves---has nevertheless remained conspicuously absent from this catalogue, despite decades of progress in atom optics and matter-wave interferometry \cite{matsukevichBell2008,kitzingerBell2021,schmiedBell2016,shinBell2019}. 

Addressing this longstanding experimental challenge is important because the motion of a massive particle is fundamentally different from an internal qubit-like state. Rather than acting as an internal quantum label, a motional state governs how matter propagates through space and responds to accelerations, rotations, and gravitational fields. Demonstrating quantum nonlocality in such states therefore extends Bell tests from internal quantum variables to the external dynamical degree of freedom of massive particles. Such tests establish the matter-wave analogue of the pioneering path- and momentum-entanglement Bell experiments that helped shape modern quantum optics \cite{rarityExperimental1990,ouweneel1988experimental}, and open new opportunities for precision atom interferometry \cite{Gross2010,Cassens_Klempt_2025} and future tests of gravity with massive quantum systems \cite{penroseGravitys1996,alvarezQuantum1989}.

Realizing such a test, as demonstrated here, has posed a long-standing experimental challenge, as it required combining several experimental capabilities that have only recently become available. A matter-wave Bell test requires the coherent generation of momentum-entangled atom pairs, high-fidelity atomic mirrors and beam splitters, phase-stable interferometry, and single-particle detection with sufficient three-dimensional momentum resolution to reconstruct the relevant correlations. Momentum-entangled pair generation, coherent Bragg manipulation, and momentum-resolved atom detection have each been demonstrated \cite{dussarratTwoParticle2017,thomasMatterwave2022,leprinceCoherent2025}, and Bell correlations in motional states have been observed \cite{athreyaBell2026}. The decisive remaining challenge was the independent implementation of local measurement bases for the two spatially separated momentum modes. Conventional Bragg lattices address resonant momentum modes globally and therefore impose a common interferometric phase. As a result, a single Bragg lattice cannot realise the independently selectable local measurement bases (local phases) required for a Bell inequality test in motional states.

In this work we overcome this limitation using a dual-resonant Bragg interferometer that enables independent control of the local phases for spatially separated momentum modes. We thereby realize the first matter-wave Bell test in which the nonlocal observables are encoded entirely in the motional states of massive particles. 
Using momentum-entangled pairs of ultracold metastable helium atoms manipulated by coherent matter-wave interferometry, we measure a Clauser–Horne–Shimony–Holt (CHSH) Bell parameter \cite{Clauser1969} of $S=2.52\pm0.17$, violating the local realistic bound by three standard deviations.

\begin{figure*}
    \centering
    \includegraphics[width=\textwidth]{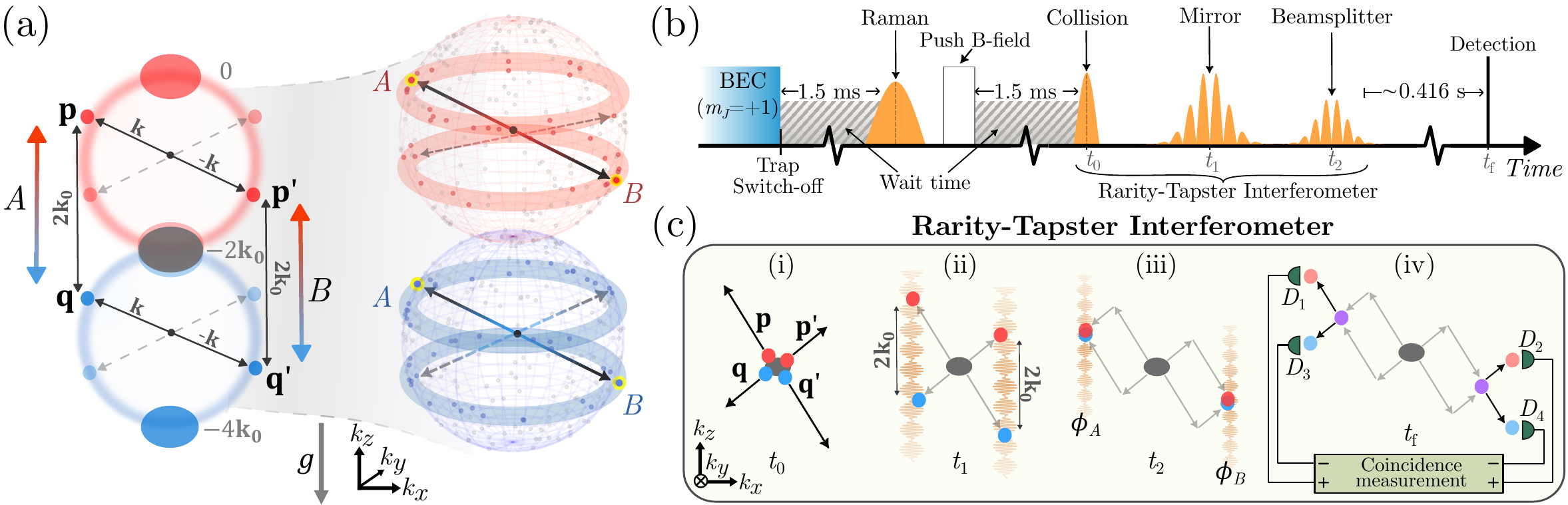}
    \caption{\textbf{Experimental schematic and Rarity-Tapster atom interferometer sequence.} (a) A Bragg collision pulse at $t_0$ coherently splits the $^4\mathrm{He}^*$ ($m_J=0$) condensate into momentum orders 0,  $-2\hbar\textbf{k}_0$, and  $-4\hbar\textbf{k}_0$. The 0 and $-2\hbar\textbf{k}_0$ components (red), and the $-2\hbar\textbf{k}_0$ and  $-4\hbar\textbf{k}_0$ components (blue), collide to produce two spherical $s$-wave scattering halos of entangled atom pairs. Within each halo, region $A$ denotes the torus-shaped region above the halo equator and contains the modes $\{\textbf{p}\}$ and $\{\textbf{q}\}$, while region $B$ denotes the corresponding torus-shaped region below the halo equator and contains the modes $\{\textbf{p}^\prime\}$ and $\{\textbf{q}^\prime\}$. The pairs ($\textbf{p},\textbf{p}^\prime$) and ($\textbf{q},\textbf{q}^\prime$) are entangled through momentum conservation of the underlying collision process. (b) Experimental timing sequence. A BEC in the $m_J=1$ sublevel is released from a magnetic trap and transferred to the $m_J=0$ sublevel by a Raman pulse, before a magnetic-field gradient pushes away the remaining $m_J=1$ atoms. After allowing the magnetic field to stabilise, a Bragg pulse at $t_0$ generates entangled atom pairs through $s$-wave collisions. The Rarity-Tapster atom interferometer sequence comprises a mirror pulse at $t_1$ and a beamsplitter pulse at $t_2$, with $t_1-t_0=t_2-t_1=350\,\mu\mathrm{s}$. Following a free-fall for $t_\text{f}\approx0.416\,\mathrm{s}$, the scattered atoms arrive at a multichannel plate and delay-line detector (MCP-DLD). (c) Schematic of the Rarity-Tapster interferometer. (i) The entangled momentum modes $\{\textbf{p},\textbf{p}^\prime,\textbf{q},\textbf{q}^\prime\}$ propagate from the $-2\hbar\textbf{k}_0$ center-of-mass momentum component. (ii) A Bragg mirror pulse transfers a momentum of $2\hbar\text{k}_0$, redirecting each mode toward its interferometric partner. (iii) A Bragg beam-splitter pulse is applied when the momentum modes $\{\textbf{p},\textbf{q}\}$ and $\{\textbf{p}^\prime,\textbf{q}^\prime\}$ overlap, coherently mixing the modes while imprinting the Bragg-beam phases $\phi_A$ and $\phi_B$ onto the two halves of the atomic superposition. (iv) The four output ports of the interferometer, $D_{\{1,2,3,4\}}$, are detected on the MCP-DLD and measure the two-particle momentum correlations between the spatially separated regions $A$ and $B$.
    }
    \label{fig:schematic}
\end{figure*}
The generation of momentum-entangled atomic pairs in our experiment begins with the preparation of a Bose–Einstein condensate (BEC) containing approximately $\sim$10$^5$ $^4$He$^*$ atoms magnetically trapped \cite{SOM} in the long-lived $2^{3}$S$_{1}$ metastable state \cite{vassenCold2012,hodgmanMetastable2009}. A rapid switch-off of the trap allows the BEC to free-fall under gravity in a stable uniform magnetic field $\textbf{B}_0\approx[0.5(\hat{\mathbf{x}}+\hat{\mathbf{z}})-0.8\hat{\mathbf{y}}]$ G, which is maintained throughout the experimental sequence (Fig.\,\ref{fig:schematic}(b)). Atomic momentum and internal-state manipulation are achieved using two orthogonal laser beams propagating along the $(\hat{\mathbf{x}}\pm \hat{\mathbf{z}})/\sqrt{2}$ direction, which drive resonant two-photon Raman and Bragg transitions \cite{thomasMatterwave2022,athreyaBell2026, SOM}. To suppress momentum distortions arising from residual magnetic-field gradients during free fall, a Raman–Bragg pulse is first applied. This pulse transfers more than 80$\%$ of the condensate population into the magnetically insensitive $m_J=0$ Zeeman sublevel while simultaneously imparting a two-photon recoil momentum of  $-2\hbar\textbf{k}_0$ in the $\hat{z}$-direction. Here $\textbf{k}_0=k_0\hat{\mathbf{z}}$, ${k}_0 = K/\sqrt{2}$ based on our beam geometry, $K=2\pi/\lambda$ is the wavenumber of the laser beam and $\lambda$=1083.19 nm denotes the wavelength of the incident laser beams.

Subsequently, a Bragg diffraction pulse \cite{khakimovGhost2016,athreyaBell2026} is applied to the moving $m_J=0$ condensate. The pulse couples momentum states separated by  $+ 2\hbar\textbf{k}_0$ and $-2\hbar\textbf{k}_0$, resulting in the coherent splitting of the condensate into three distinct momentum components with momenta 0,  $-2\hbar\textbf{k}_0$, and  $-4\hbar\textbf{k}_0$.  These momentum components contain approximately 25\%, 50\%, and 25\% of the atomic population, respectively. As these momentum-diffracted condensate packets propagate through each other, spontaneous $s$-wave scattering events \cite{perrinObservation2007,perrinAtomic2008} between momentum modes (0,$-2\hbar\textbf{k}_0$) and ($-2\hbar\textbf{k}_0$,$-4\hbar\textbf{k}_0$) generate pairs of atoms with equal and opposite momenta, forming two spatially separated spherical scattering halos in momentum space (Fig.\,\ref{fig:schematic}(a)). 

The collision process conserves both momentum and energy, leading to the creation of momentum-entangled atom pairs occupying diametrically opposite modes on the scattering halo \cite{athreyaBell2026}. In the double-halo configuration, coherent scattering occurs simultaneously into two spatially separated halos, providing two independent sources of correlated atom pairs \cite{thomasMatterwave2022}. By operating in the low-gain regime, where the mean occupation number per scattered momentum mode satisfies $\bar{n}\ll1$, the probability of generating multiple atom pairs within the same mode is strongly suppressed \cite{hodgmanSolving2017}. In our experiment, we estimate $\bar{n}$ to be on the order of $\sim 0.02$ \cite{SOM}. Under these conditions, the scattering process produces a set of well-defined two-particle momentum-entangled states of the form \cite{lewis-swanProposal2015a,thomasMatterwave2022}:
\begin{equation}
\label{eq:bell}
\ket{\Psi} \approx \frac{1}{\sqrt{2}}(\ket{1}_{\textbf{p}}\ket{1}_{\textbf{p}^\prime}\ket{0}_{\textbf{q}}\ket{0}_{\textbf{q}^\prime} + \ket{0}_{\textbf{p}}\ket{0}_{\textbf{p}^\prime}\ket{1}_{\textbf{q}}\ket{1}_{\textbf{q}^\prime}),
\end{equation}
where (\textbf{p},$\textbf{p}^\prime$) and (\textbf{q},$\textbf{q}^\prime$) correspond to correlated momentum modes in the top (red) and bottom (blue) halos, respectively (as illustrated in Fig.\,\ref{fig:schematic}(a)), and satisfying, $\textbf{p}+\textbf{p}^\prime=-2\textbf{k}_0$,  $\textbf{q}+\textbf{q}^\prime=-6\textbf{k}_0$, 
$\textbf{p}-2\textbf{k}_0 =\textbf{q}$ and $\textbf{p}^\prime-2\textbf{k}_0 =\textbf{q}^\prime$. Equation \eqref{eq:bell} represents a coherent superposition of two indistinguishable pair-production pathways and is directly analogous to the maximally entangled Bell state $\frac{1}{\sqrt{2}}(|HH\rangle+|VV\rangle)$, in the horizontal/vertical ($H/V$) polarization basis \cite{thomasMatterwave2022,lewis-swanProposal2015a, kitagawaPhaseSensitive2011}.

Atom detection is performed using a microchannel plate and delay-line detector system \cite{kannanMeasurement2024,manningHanbury2010}, enabling reconstruction of the full three-dimensional momentum of each atom and measurement of momentum-space correlations \cite{hodgmanSolving2017}. Characterisation of the double-halo source via the second-order back-to-back correlation function \cite{hodgmanSolving2017,athreyaBell2026} reveals strong intra-halo correlations and negligible cross-halo correlations, confirming spontaneous four-wave mixing \cite{perrinObservation2007} while suppressing contributions from higher-order scattering processes. These observations validate the two-particle Bell-state approximation of Eq.\,\eqref{eq:bell} \cite{lewis-swanProposal2015a,SOM}.

\begin{figure}
    \centering
\includegraphics[width=0.46\textwidth]{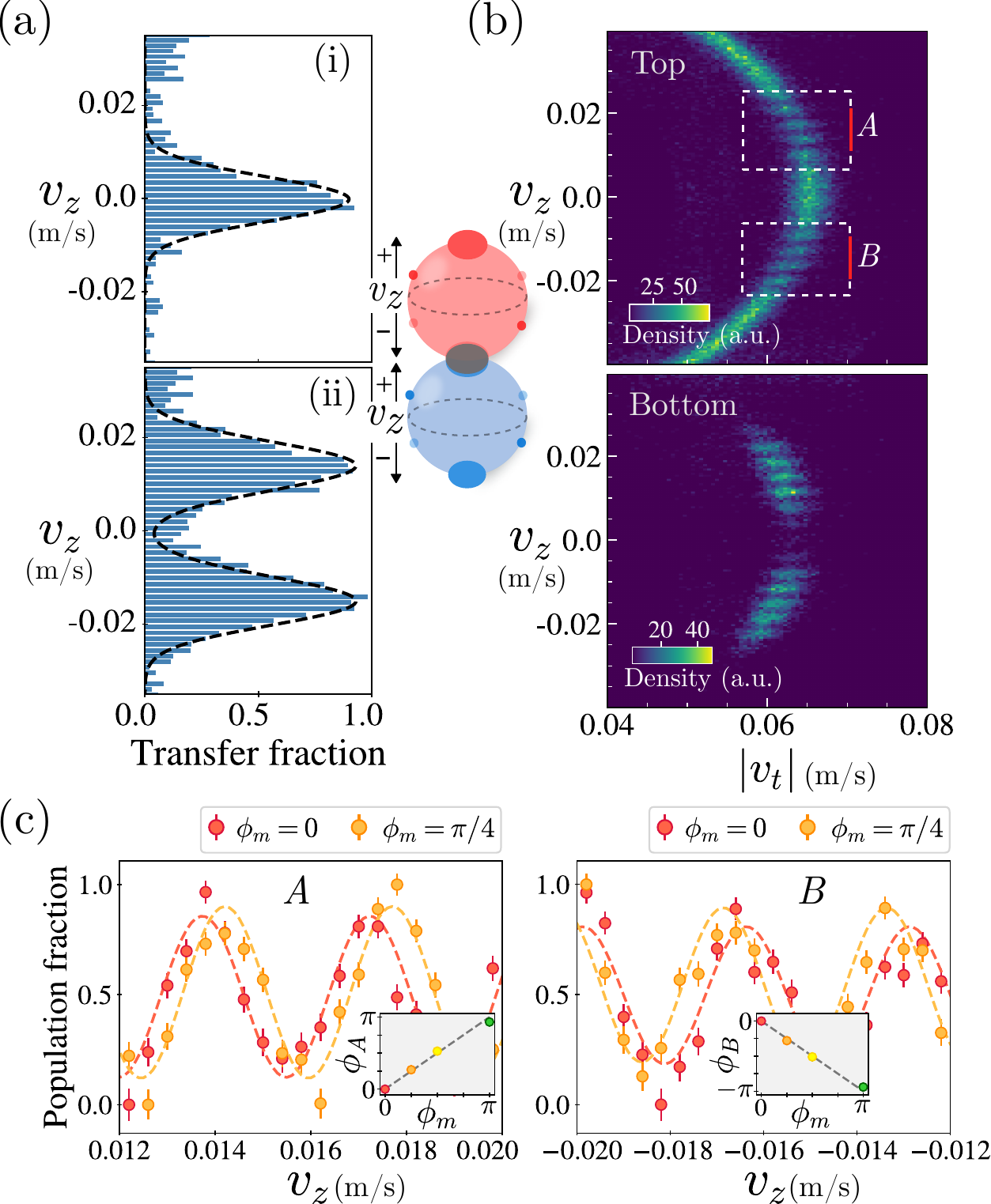}
    \caption{\textbf{Characterization of the dual-resonant Bragg pulse.} (a) Transfer fraction as a function of vertical velocity $v_z$. (i) For $\omega_m =0$, the Bragg pulse is a standard single-frequency pulse with a single resonance at the halo equator ($v_z=0$). (ii) For $\omega_m=2\pi\times21\,\mathrm{kHz}$, the dual-resonant Bragg pulse exhibits two resonances at $v_z=\pm0.016\,\mathrm{m}\, \mathrm{s}^{-1}$, corresponding to the off-equatorial halo regions $A$ and $B$. The peak transfer efficiency is $92(4)\%$. (b) Two-dimensional histogram of the Ramsey-like interferometer \cite{SOM} output, plotted as vertical velocity $v_z$ versus transverse velocity $v_t$ (integrated over the $x$-$y$ plane) for the `Top' and `Bottom' halos. Interference fringes are observed in the two spatially separated halo regions $A$ and $B$ dependent on the imprinted phase $\phi_m$. (c) Population fraction as a function of $v_z$ in regions $A$ and $B$ for modulation phases $\phi_m=0$ (red) and $\phi_m=\pi/4$ (orange). The fringe phases $\phi_A$ and $\phi_B$ are extracted from sinusoidal fits. Insets: extracted phases $\phi_A$ and $\phi_B$ as functions of $\phi_m$. Error bars represent the standard deviation of the histogrammed bin counts.}
    \label{fig:pulse}
\end{figure}

Previous studies employed the double-halo geometry to demonstrate Bell correlations between momentum-entangled atom pairs using a Rarity–Tapster-type interferometer \cite{athreyaBell2026,yanProposal2025}. As discussed in the introduction, these implementations relied on conventional single-frequency Bragg lattices, which couple all resonant momentum modes globally and therefore preclude independently selectable local measurement bases. Here we overcome this limitation using a dual-resonant Bragg lattice \cite{leprinceCoherent2025,lellouchPolychromatic2023,louieRobust2023}. By selectively addressing distinct momentum classes within the two halves of the scattering halos and imprinting independently tunable interferometric phases on these halves (illustrated in Fig.~\ref{fig:schematic}(c)), the local measurement settings required for a CHSH Bell test can be realized \cite{clauserProposed1969,clauserBells1978,lewis-swanProposal2015a}.

The dual-resonant lattice is generated through amplitude modulation of the two-photon Bragg coupling. Unlike a conventional single-frequency Bragg pulse, which uniformly addresses all resonant momentum modes, the modulated lattice produces two phase-coherent frequency components that can be tuned to select only distinct classes of momentum states within the scattering halos.
The effective two-photon Rabi frequency is given by:
\begin{equation}
\label{eq:rabi}
\Omega_R(t) = \Omega(t)\cos(\omega_m t + \phi_m),
\end{equation}
where $\Omega(t)=\Omega_0\exp[-t^2/(2\sigma^{2})]$ is a Gaussian pulse envelope of amplitude $\Omega_0$ and temporal width $\sigma$, $\omega_m$ is the modulation frequency, and $\phi_m$ is the modulation phase. In the frequency domain, this modulation generates two phase-coherent sidebands at frequencies $$\omega_{A,B} = \omega_0 \pm \omega_m,$$ where $\omega_0$ denotes the two-photon frequency difference between the Bragg laser beams. For sufficiently long pulse durations, these sidebands correspond to two Bragg resonances with effective detunings shifted by $\pm\omega_m$ relative to the carrier resonance $\omega_0$. By appropriately choosing $\omega_m$, the two sidebands can be brought into resonance with distinct momentum classes within the scattering halo, allowing a single optical pulse to simultaneously drive two independent Bragg transitions.

This spectral selectivity arises from the momentum dependence of the Bragg resonance condition. The Doppler-shifted two-photon detuning is given by $\delta(\textbf{v}) = \omega_0 - 2\textbf{k}_0\cdot\textbf{v}$, where $\textbf{v}$ is the velocity of the atom. Since the Doppler shift depends on the projection of the atomic momentum along the lattice direction, the Bragg resonance frequency varies across the scattering halo. In particular, atoms occupying off-equatorial regions experience distinct resonance frequencies, enabling different momentum classes to be spectrally resolved. By selecting the modulation frequency appropriately, the two sidebands $\omega_A$ and $\omega_B$ become resonant with atoms propagating toward opposite regions of the halo. Consequently, each atom in an entangled pair is coupled by different frequency components of the same optical lattice. In our implementation, atoms occupying momentum modes ($\textbf{p}$,$\textbf{q}$) are resonantly driven by $\omega_A$, whereas atoms occupying modes ($\textbf{p}^\prime$,$\textbf{q}^\prime$) are resonantly driven by $\omega_B$ (see Fig.\ref{fig:schematic}(a)).
The phase imprinted by each Bragg transition is determined by both the optical phase difference between the Bragg laser beams and the modulation phase. Denoting the optical phase difference by $\phi_0$ and modulation phase by $\phi_m$, the two sidebands drive transitions with effective interferometric phases
\begin{equation}
\label{eq:phases_AB}
\phi_A=\phi_0+\phi_m,
\qquad\
\phi_B=\phi_0-\phi_m.
\end{equation}

As a result, atoms addressed by the two resonances acquire distinct interferometric phases despite interacting with the same optical lattice. Thus, $\phi_m$ provides direct control over the relative measurement basis while referencing a common optical phase $\phi_0$ between the two Bragg transitions. The independent variations of phases $\phi_A$ and $\phi_B$ thus constitute the independently selectable local measurement settings
required for the implementation of a CHSH Bell test.

The spectral selectivity and transfer efficiency of the dual-resonant Bragg pulses are characterised using a single scattering halo \cite{SOM}. Figure \ref{fig:pulse}(a) shows the momentum transfer fraction for Bragg coupling between the momentum modes ($\mathbf{p},\mathbf{p}'$) and ($\mathbf{q},\mathbf{q}'$). Figure \ref{fig:pulse}(a, i) corresponds to an unmodulated pulse ($\omega_m=0$), equivalent to a conventional single-frequency Bragg pulse, which couples only a single resonant pair of momentum modes \cite{thomasMatterwave2022,athreyaBell2026}. Figure \ref{fig:pulse}(a, ii) corresponds to a modulated pulse ($\omega_m=2\pi\times21\,\mathrm{kHz}$), where coupling is restricted to off-equatorial modes displaced by $\sim\pm0.016\,\mathrm{m/s}$ from the halo equator, demonstrating the spectral selectivity of the dual-resonant pulse.

We verify that we can achieve independent phase control through Ramsey-like interferometry \cite{ramseyMolecular1950,leprinceCoherent2025,SOM}. Using dual-resonant Bragg pulses, we realise two parallel interferometers within a single scattering halo, corresponding to the mode-selected regions $A$ and $B$ addressed by frequency components $\omega_A$ and $\omega_B$, respectively. The resulting interference fringes are shown in Fig.\,\ref{fig:pulse}(b). The extracted interferometric phases exhibit a linear dependence on the applied modulation phase, with $\phi_A \!\propto\! \phi_m$ and $\phi_B \!\propto \!-\phi_m$, as shown in Fig.\,\ref{fig:pulse}(c) \cite{SOM}.

\begin{figure}
    \centering
    \includegraphics[width=0.47\textwidth]{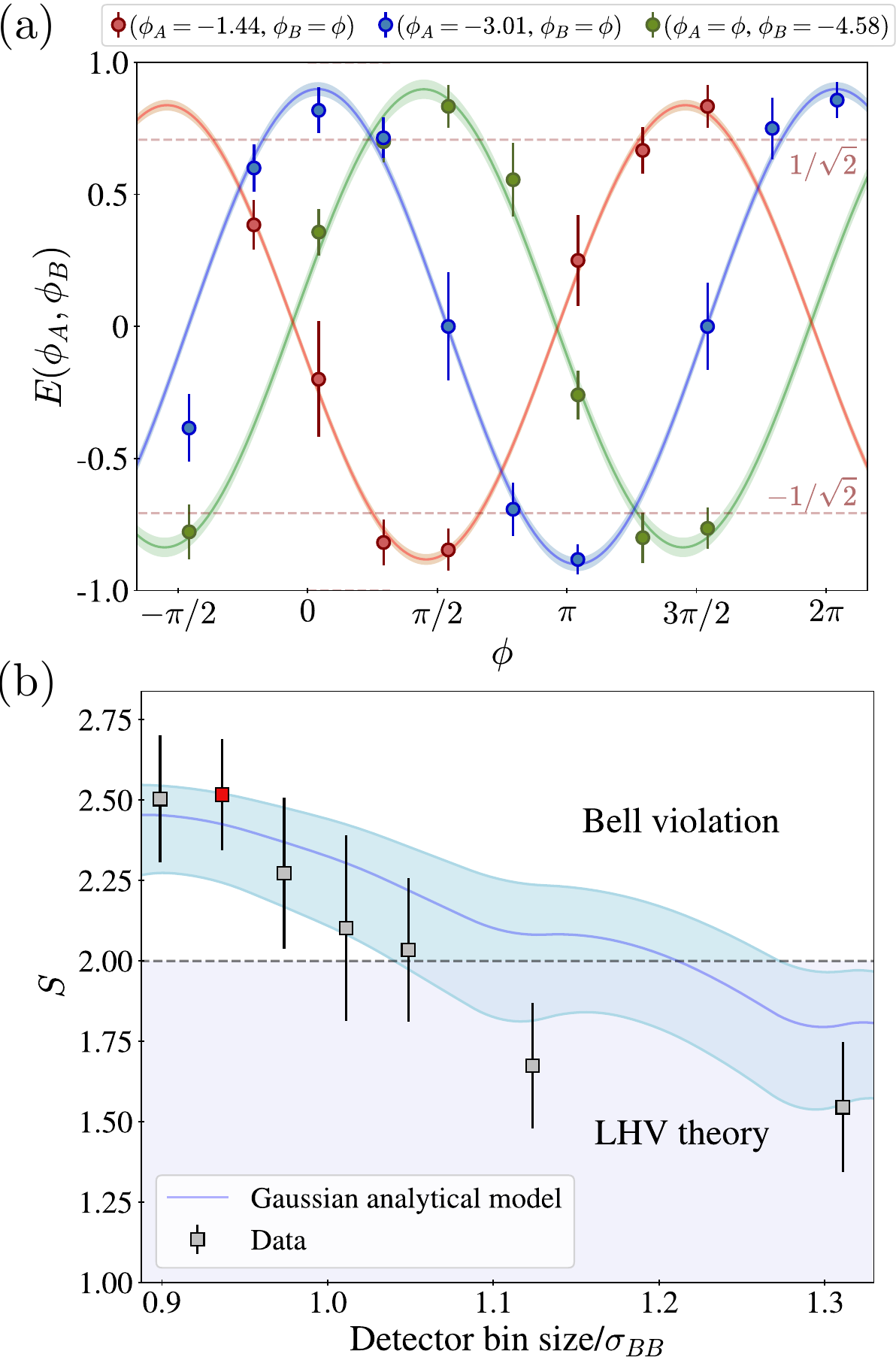}
    \caption{\textbf{Bell correlation measurements and CHSH-Bell parameter.} (a) Measured Bell correlation function $E(\phi_A,\phi_B)$ for three phase configurations. The red, blue, and green data sets correspond to $(\phi_A,\phi_B)=(-1.44,\phi)$, $(-3.01,\phi)$, and $(\phi,-4.58)$, respectively. The data are fitted with sinusoidal functions of the form given in Eq.\,\eqref{eq:Ecorr_theory}, yielding an average visibility of $V=0.93(2)$. Error bars are estimated using a binomial proportion estimator \cite{Cetinkaya-RundelOpenIntroStat2019}. The data are obtained from 38,300 experimental shots. (b) CHSH-Bell parameter $S$ as a function of detector bin size, in units of the back-to-back momentum correlation length $\sigma_{\mathrm{BB}}$. Grey points show the measured values, with the red point indicating the maximum nonlocality violation $S=2.52\pm0.17$. Error bars show the sinusoidal fit uncertainties of the Bell correlation functions. The purple line shows the Gaussian analytical model of Ref.~\cite{lewis-swanProposal2015a,thomasMatterwave2022} using experimental parameters. The blue shaded region shows the uncertainty of the model from the measured correlation amplitudes.}
    \label{fig:Ecorr_S}
\end{figure}

Having established independent control of the measurement settings, we implement a Rarity–Tapster-type matter-wave interferometer \cite{thomasMatterwave2022,rarityExperimental1990} to probe the nonlocal correlations of the momentum-entangled atom pairs. Following the collision pulse at time $t_0$, dual-resonant Bragg mirror and beam-splitter pulses are applied at $t_1=t_0+350\,\mu\mathrm{s}$ and $t_2=t_1+350\,\mu\mathrm{s}$, respectively, as illustrated in Fig.\,\ref{fig:schematic}(b, c). The independently controlled phases $\phi_A$ and $\phi_B$ of the beam splitter pulse define the measurement settings for the two atoms occupying the selected off-equator momentum modes. The interferometer coherently recombines the two pair-production pathways associated with Eq.\,\eqref{eq:bell}, causing the joint atom detection probabilities at the output ports $D_{\{1,2,3,4\}}$ (Fig.\,\ref{fig:schematic}(c)) to oscillate as a function of the interferometric phases \cite{SOM}. 

We quantify the joint atom detection probabilities or atom-atom correlations at the output ports $D_{\{1,2,3,4\}}$  through the four distinct joint probability distributions $P_{\textbf{k},\textbf{k}^\prime}$, where $\textbf{k}\in\{\textbf{p},\textbf{q}\}$ and $\textbf{k}^\prime\in\{\textbf{p}^\prime,\textbf{q}^\prime\}$.
From these probabilities we construct the Bell correlation function $E(\phi_A,\phi_B)$ \cite{clauserBells1978}, 
\begin{equation}
\label{eq:Ecorr}
    E(\phi_A,\phi_B) = \frac{P_{\textbf{p},\textbf{p}^\prime}+P_{\textbf{q},\textbf{q}^\prime}-P_{\textbf{p},\textbf{q}^\prime}-P_{\textbf{q},\textbf{p}^\prime}}{P_{\textbf{p},\textbf{p}^\prime}+P_{\textbf{q},\textbf{q}^\prime}+P_{\textbf{p},\textbf{q}^\prime}+P_{\textbf{q},\textbf{p}^\prime}},
\end{equation}
which takes the general form \cite{clauserBells1978}
\begin{equation}
    \label{eq:Ecorr_theory}
    E(\phi_A,\phi_B) = -V \cos(\phi_A+\phi_B),
\end{equation}
where $V$ is the interference visibility.

To display the cosine phase dependence of $E$---the hallmark of coherent two-particle interference---we perform complementary measurements in which one interferometer phase is held fixed while the other is varied. Figure \ref{fig:Ecorr_S}(a) shows the measured correlation function for several combinations of $\phi_A$ and $\phi_B$. Fitting the data to Eq.\,\eqref{eq:Ecorr_theory} yields an average visibility of $V=0.93(2)$, consistent across all measurement configurations and in good agreement with the expected visibility of $V=0.95$ obtained from a model incorporating higher-order Fock-state contributions in the entangled pair generation process and path-phase averaging over the finite detection-port integration windows \cite{SOM}. The reduction from the ideal value ($V=1$) is therefore well accounted for by these experimental imperfections.

From these measurements, we extract $E$ for four pairs of phase   settings $\phi_A\!\in\!\{\phi_{A}^{(1)},\phi_{A}^{(2)}\}$ and $\phi_B\!\in\!\{\phi_{B}^{(1)},\phi_{B}^{(2)}\}$, for evaluation of the CHSH-Bell parameter \cite{rarityExperimental1990,clauserProposed1969}
\begin{align}
S\! =&\!\left|E(\phi_{A}^{(1)},\phi_{B}^{(1)})\!-\!E(\phi_{A}^{(1)},\phi_{B}^{(2)})\!\right. \nonumber\\
&+\left.E(\phi_{A}^{(2)},\phi_{B}^{(1)})\!+\!E(\phi_{A}^{(2)},\phi_{B}^{(2)})\right|,
    \label{eq:S}
\end{align}
for which all local hidden variable (LHV) theories satisfy $S\leq2$. The four phase settings and the results of the Bell correlation measurements obtained with them are given in Table \ref{tab:Ecorr}, based on approximately 11,500 experimental shots. Each term is determined from independently acquired datasets under identical (apart from $\phi_{A,B}$) experimental conditions, ensuring statistical independence between the different measurement settings. The uncertainties in $E$ are estimated using a binomial proportional estimator \cite{Cetinkaya-RundelOpenIntroStat2019,SOM} and propagated to obtain the uncertainty in $S$. Combining these data through Eq.~\eqref{eq:S}, we obtain 
\begin{equation}
S=2.52\pm0.17,
\end{equation}
corresponding to a violation of the CHSH-Bell inequality by approximately three standard deviations and demonstrating nonlocal correlations between momentum-entangled pairs of massive particles.

To investigate the robustness of the violation, we further examine the dependence of the extracted Bell parameter on the detector integration volume. Figure \ref{fig:Ecorr_S}(b) shows the measured CHSH-Bell parameter $S$ as a function of the detection-bin size, expressed in units of the back-to-back momentum-correlation length $\sigma_{\mathrm{BB}}$ \cite{ogrenAtomatom2009,hodgmanSolving2017}. The measured values are consistent, within uncertainties, with the Gaussian analytical model developed in \cite{thomasMatterwave2022,lewis-swanProposal2015a}, validating its applicability as a predictive tool for pair-correlated scattering experiments in ultracold atom interferometers.

\begin{center}
    \begin{table}[]
        \begin{tabular}{C{2cm} C{1.5cm} c}
        \hline\hline
        \\[-2.2ex]
        $\phi_A$& $\phi_B$ & $E(\phi_A,\phi_B)$\\ [0.5ex]
        \hline
            $-1.436$ & $-0.652$ & $0.385\pm0.093$ \\
            $-1.436$ & $0.919$ & $-0.818\pm0.076$ \\
            $-3.007$ & $-0.652$ & $0.601\pm0.087$ \\
            $-3.007$ & $0.919$ & $0.714\pm0.089$ \\
        \hline\hline
        \end{tabular}
        \caption{\textbf{Measurement phase settings for the CHSH violation}. The four phase combinations $(\phi_A,\phi_B)$ and corresponding Bell correlation values $E(\phi_A,\phi_B)$ extracted from Fig.\,\ref{fig:Ecorr_S}(a) are used to determine the CHSH-Bell parameter $S$. Uncertainties in $E$ are from the binomial proportion estimator.}
    \label{tab:Ecorr}
    \end{table}
\end{center}

Our results establish the combination of the double-halo momentum-entangled source and dual-resonant Bragg control as a powerful platform for momentum-space quantum-atom optics. The double-halo geometry provides a scalable source of momentum-entangled atom pairs, while the dual-resonant Bragg lattice enables independent and coherent manipulation of relative phases of distinct pairs of well-separated momentum modes. Together, these capabilities provide the essential resources required for controlled investigations of nonlocal quantum correlations in systems of massive particles.

Beyond Bell tests, the independent phase control enabled by the dual-resonant Bragg lattice supports measurements in arbitrary interferometric bases, providing a route to full quantum-state tomography \cite{maurodarianoQuantum2003} and comprehensive characterization of entanglement and other non-classical properties. Together with the momentum-entangled double-halo source, this establishes a versatile platform for studies of quantum nonlocality \cite{brunnerBell2014}, multipartite entanglement \cite{amicoEntanglement2008}, and quantum-information protocols \cite{braunsteinQuantum2005} with motional degrees of freedom of massive particles. Looking forward, the platform also enables matter-wave Bell tests involving momentum-entangled isotopes, particularly $^3$He$^*$ and $^4$He$^*$ \cite{yanProposal2025}. This would provide a unique testbed for exploring the interplay between quantum nonlocality, gravity, and massive-particle entanglement \cite{penroseGravitys1996,alvarezQuantum1989}.


\bibliography{CHSH_bell}


\section*{Acknowledgments}
The authors would like to thank A. Aspect, K.F. Thomas and P. Treutlein for insightful discussions. This work was supported through the Australian Research Council (ARC) Discovery Projects, Grant Nos. DP190103021, DP240101346, DP240101441 and DP240101033. S.S.H. was supported by the Australian Research Council Future Fellowship Grant No. FT220100670. S.K. was supported by an Australian Government Research Training Program scholarship.

\clearpage
\onecolumngrid
\begin{center}
\textbf{{\mdseries\large Supplementary material for}\\ \large Bell inequality violation with momentum-entangled massive particles}
\vspace{1cm}
\end{center}

\setcounter{equation}{0}
\setcounter{figure}{0}
\setcounter{table}{0}
\setcounter{section}{0}
\renewcommand{\figurename}{\textbf{Fig.}}
\renewcommand{\thefigure}{\textbf{S}\textbf{\arabic{figure}}}
\renewcommand{\theequation}{\text{S.}\arabic{equation}}

\section{\label{Experiment} Experimental apparatus and atom-optical control}
We prepare a Bose–Einstein condensate (BEC) of metastable $^4$He$^*$ atoms in a harmonic magnetic trap formed by a bi-planar quadrupole-Ioffe configuration \cite{dallBose2007}, with characteristic trapping frequencies $(\omega_x,\omega_y,\omega_z)/2\pi = (15,25,25)\,\mathrm{Hz}$. The vertical $\hat{z}$-direction defines the collision axis for the subsequent $s$-wave scattering process \cite{perrinObservation2007,perrinAtomic2008} and the direction of Bragg momentum transfer.

The coherent manipulation of the atomic internal and momentum states is achieved using a combination of Raman and Bragg transitions \cite{athreyaBell2026,thomasMatterwave2022}. Two orthogonal laser beams, propagating along the $(\hat{\mathbf{x}}\pm \hat{\mathbf{z}})/\sqrt{2}$ directions, are frequency locked and blue-detuned by $2.3\,\mathrm{GHz}$ from the $2^{3}$S$_{1}-$$2^3$P$_0$ transition (Fig.\,\ref{Ramsey scheme_fig}(a)), thereby suppressing unwanted single-photon excitation. Sequences of Raman and Bragg pulses are used to coherently transfer populations between selected internal and momentum states, as described in the main text and illustrated in the experimental timing sequence (Fig.\,1(b)).

A Gaussian amplitude-modulated Raman pulse with duration $10\,\mathrm{\mu s}$ transfers population from the $m_J=+1$ to the $m_J=0$ internal state while imparting a momentum transfer of $-2\hbar\textbf{k}_0$. A subsequent Bragg diffraction pulse with duration $5\,\mathrm{\mu s}$ couples momentum states (0,  $-2\hbar\textbf{k}_0$, and  $-4\hbar\textbf{k}_0$) within the $m_J=0$ manifold during the collision process. These atom-optical control sequences prepare and manipulate the momentum modes forming the input states for the dual-resonant Bragg interferometer described below.

\section{\label{DRB} Dual-resonant Bragg Interferometery and Characterisation}
\subsection{Dual-resonant Bragg lattice and momentum-mode addressing}
The optical potential generated by a conventional single-frequency Bragg lattice \cite{meystre2001atom} is given by
\begin{equation}
    V(x,t)=V_0 (t)\cos(\Delta kx-\omega_0 t + \phi_0)
\end{equation}
where $V_0(t)$ is the pulse envelope, $\Delta k = |\Delta\textbf{k}| =2k_0$ is the momentum transfer, $\omega_0$ is the two-photon frequency difference between the Bragg beams, and $\phi_0$ is their relative optical phase. This potential couples momentum states satisfying the Bragg resonance condition \cite{kozumaCoherent1999},
\begin{equation}
    \omega=\frac{\hbar(\Delta k)^2}{2m} + \frac{\hbar}{m}\textbf{k}\cdot\Delta\textbf{k}.
\end{equation}
In contrast to previous implementations of our matter-wave Rarity–Tapster interferometer \cite{thomasMatterwave2022,athreyaBell2026}, here we implement a dual-resonant Bragg lattice \cite{leprinceCoherent2025,lellouchPolychromatic2023,louieRobust2023}, which simultaneously addresses two distinct pairs of momentum modes in the scattering halo. This is achieved by applying a sinusoidal amplitude modulation to the Bragg lattice,
\begin{equation}
    V(x,t)=V_0 (t)\cos(\Delta kx-\omega_0 t + \phi_0)\cos(\omega_mt+\phi_m).
\end{equation}
Using the trigonometric identity $\cos(a)\cos(b)=\frac{1}{2}[\cos(a+b)\cos(a-b)]$, this can be written as
\begin{equation}
    V(x,t)=\frac{V_0(t)}{2}[\cos(\Delta kx-(\omega_0+\omega_m)t+(\phi_0+\phi_m))+\cos(\Delta kx-(\omega_0-\omega_m)t+(\phi_0-\phi_m))].
\end{equation}
Thus, a single amplitude-modulated pulse generates two frequency components symmetrically displaced around the carrier frequency, $\omega_0\pm\omega_m$. These components are independently resonant with the Doppler-shifted momentum classes located in regions $A$ and $B$ of the halo. The modulation phase therefore produces opposite phase shifts in the two resonant channels, providing independent control of the interferometric phases while maintaining an identical pulse envelope.

Equivalently, the dual-resonant lattice can be described as two phase-controlled optical fields,
\begin{equation}
    E_A(t) \propto\,e^{-i(\omega_At+\phi_A)},\qquad E_B(t) \propto\,e^{-i(\omega_Bt+\phi_B)},
\end{equation}
where
\begin{equation}
    \omega_A=\omega_0+\omega_m,\qquad \omega_B=\omega_0-\omega_m, 
\end{equation}
and
\begin{equation}
    \phi_A=\phi_0+\phi_m,\qquad \phi_B=\phi_0-\phi_m.
\end{equation}

In our experiment, the modulation frequency is set to $\omega_m=2\pi\times21\,\mathrm{kHz}$, which separates the two resonantly coupled momentum classes by approximately $0.032\,\text{m/s}$, corresponding to offsets of $\pm0.016\,\text{m/s}$ from the halo equator. For pulse durations satisfying $\tau\gg \pi/\omega_m$, the sideband separation exceeds the Fourier-limited spectral width, enabling independent coupling of the two momentum classes with negligible overlap.
\begin{figure}
    \centering
    \includegraphics[width=0.9\linewidth]{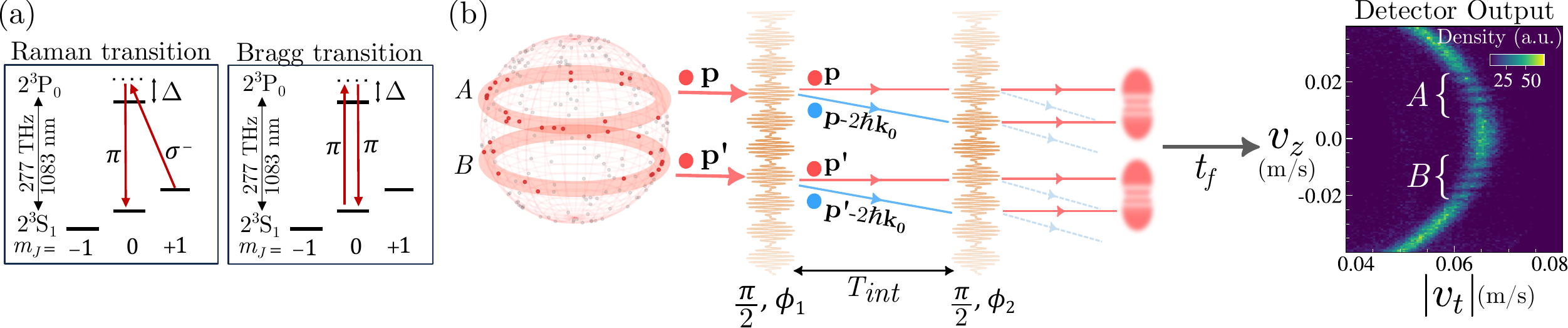}
    \caption{\textbf{Raman-Bragg transitions and interferometer phase calibration.} (a) Raman and dual-resonant Bragg two-photon transitions used in the experiment. The lasers are blue-detuned by $\Delta = 2.3~\mathrm{GHz}$ from the $2^3\text{S}_1 \rightarrow 2^3\text{P}_0$ transition. (b) Schematic of the Ramsey-type sequence used to calibrate the interferometer phase imparted by the dual-resonant Bragg pulses, applied to a single halo containing the $(\textbf{p},\textbf{p}^\prime)$ momentum modes. Two $\pi/2$ Bragg pulses, carrying optical phases $\phi_1$ and $\phi_2$, are separated by an interrogation time $T_{int}=250\,\mu\mathrm{s}$. The first pulse creates a 50/50 superposition of momentum states $\{\textbf{p}, \textbf{p}-2\hbar \textbf{k}_0\}$ and $\{\textbf{p}^\prime, \textbf{p}^\prime-2\hbar \textbf{k}_0\}$, while the second pulse recombines the momentum pathways. Atoms with the same final momenta ($\textbf{p}$ or $\textbf{p}^\prime$) interfere upon detection after a fall time $t_f\approx0.416\,\mathrm{s}$. The resulting detector output, shown as a 2D histogram of vertical velocity $v_z$ versus transverse velocity $v_t$, exhibits interference fringes in regions $A$ and $B$, corresponding to momentum modes $\textbf{p}$ and $\textbf{p}^\prime$, respectively.}
    \label{Ramsey scheme_fig}
\end{figure}

\subsection{Bragg transfer characterisation and interferometer phase calibration}
The dual-resonant Bragg $\pi$- and $\pi/2$-pulse operations are characterised by measuring population transfer as a function of pulse duration, amplitude, detuning, and timing \cite{athreyaBell2026}. To isolate the transfer between the coupled momentum modes, we first prepare a single halo containing either the (\textbf{p},$\textbf{p}^\prime$)-Top halo, or (\textbf{q},$\textbf{q}^\prime$)-Bottom halo, momentum modes and apply the Bragg pulse to transfer population between the counter-propagating momentum pairs, (\textbf{p},$\textbf{p}^\prime$)$\to$(\textbf{q},$\textbf{q}^\prime$) or (\textbf{q},$\textbf{q}^\prime$)$\to$(\textbf{p},$\textbf{p}^\prime$).
Optimising both the pulse duration and intensity in each operation, we obtain mirror and beamsplitter pulse durations of $\tau_\pi=75.2\,\mu\text{s}$ and $\tau_{\pi/2}=71.7\,\mu\text{s}$, respectively, with the slightly shorter $\pi/2$ operation reflecting its lower optimized intensity and thus reduced effective Rabi frequency compared to the $\pi$ pulse. The corresponding peak transfer efficiencies are 92(4)$\%$ for the mirror operation and 50(2)$\%$ for the beamsplitter operation. The measured transfer profiles are centred at $\pm 0.016\,\text{m/s}$ from the halo equator and exhibit approximately Gaussian momentum widths of $0.0025\,\text{m/s}$. The transfer peaks are equal and symmetric in the two coupled momentum regions $A$ and $B$. Figure \,2(a) shows the transfer fraction profile achieved by the optimised mirror pulse in our setup. 

The interferometric phase control is calibrated using a Ramsey-type sequence consisting of two dual-resonant $\pi/2$-pulses separated by an interrogation time $T_{int}$ \cite{ramseyMolecular1950,leprinceCoherent2025} (Fig.\,\ref{Ramsey scheme_fig}(b)). The two pulses are identical except for their optical phases. The first pulse has phase $\phi_1$, while the second pulse has phase $\phi_2$. Starting from a halo containing the (\textbf{p},$\textbf{p}^\prime$) momentum modes, the two pulses create interference fringes after a time $T_{int}=250\,\mu\text{s}$.

For the first pulse, the modulation phase is set to zero,
\begin{equation}
    \phi^A_1=\phi^B_1=\phi_0,
\end{equation}
while for the second pulse,
\begin{align}
    \phi^A_2=\phi_0 +\phi_m,\\
    \phi^B_2=\phi_0 -\phi_m. 
\end{align}
Therefore,
\begin{equation}
    \phi^A_2-\phi^A_1=\phi_m,\quad\text{and}\quad\phi^B_2-\phi^B_1=-\phi_m,
\end{equation}
resulting in opposite phase shifts of the interference fringes from regions $A$ and $B$. The final interferometric phase is
\begin{equation}
    \Phi_{Ramsey} = -\phi_1+\phi_2+\phi_{grav}+2kg t_f T_{int},
\end{equation}
where $\phi_{grav}$ accounts for the constant gravitational phase accumulated by the interferometer arms between the Bragg pulses and $t_f$ is the subsequent fall time to the detector ($\sim0.416\,\text{s}$). The resulting Ramsey fringes, obtained by averaging over 1000 experimental shots (Fig.\,\ref{Ramsey scheme_fig}(b), Detector Output), are used to extract the phase response of each momentum-selected region, demonstrating independent and stable phase control of the two dual-resonant Bragg channels.

\subsection{Interferometer model and joint detection probabilities}
Following the formalism of Refs.\,\cite{athreyaBell2026,thomasMatterwave2022}, the dual-resonant Rarity–Tapster interferometer is modelled as two independent Bragg coupling channels, each channel resonant with the selected momentum modes in regions $A$ or $B$. Since the two frequency components are spectrally resolved, each coupling channel is described by the same two-mode Bragg Hamiltonian,
\begin{equation}
    \hat{H}=\frac{\hbar\Omega}{2}
    \begin{pmatrix}
        0 & e^{i\phi}\\
        e^{-i\phi} & 0 \\
    \end{pmatrix},
\end{equation}
where $\Omega$ is the two-photon Rabi frequency and $\phi$ is the phase of the Bragg lattice. Using the unitary evolution operator $\hat{U}(t,\phi) = e^{-i\hat{H}t/\hbar}$, we model the Rarity-Tapster interferometer consisting of the $\pi$-pulse mirror ($\hat{U}(\pi/\Omega,\pi/2)$) and $\pi/2$-pulse beamsplitter ($\hat{U}(\pi/2\Omega,\phi_{A,B})$) operations.

The two resonant channels correspond to the momentum regions $A$ and $B$, with beamsplitter phases $\phi_A$ and $\phi_B$, respectively. Applying the interferometer transformation to the entangled input state (Eq.\,(1))  gives joint detection probabilities between the output momentum modes. Defining $P_{\textbf{k},
\textbf{k}^\prime}$ as the probability of detecting a correlated pair in output modes $\textbf{k}$ and $\textbf{k}^\prime$, where $\textbf{k}\in\{\textbf{p},\textbf{q}\}$ and $\textbf{k}^\prime\in\{\textbf{p}^\prime,\textbf{q}^\prime\}$, the resulting probabilities are
\begin{equation}
P_{\textbf{p},\textbf{p}^\prime}=P_{\textbf{q},\textbf{q}^\prime} = \frac{1}{2}\sin^2\left(\frac{\phi_A+\phi_B}{2}\right),\quad\text{and}\quad
P_{\textbf{p},\textbf{q}^\prime}=P_{\textbf{q},\textbf{p}^\prime} = \frac{1}{2}\cos^2\left(\frac{\phi_A+\phi_B}{2}\right).
\end{equation}
Thus, the interference depends on the sum of the two Bragg phases rather than their relative phase. This distinction arises from the slightly different, yet physically equivalent, geometry of our Rarity-Tapster interferometric scheme \cite{rarityExperimental1990,thomasMatterwave2022}.

Experimentally, the joint detection probabilities are extracted from the two-particle momentum correlation function \cite{hodgmanSolving2017,athreyaBell2026},
\begin{equation}
    g^{(2)}(\textbf{k},\textbf{k}^\prime) = \frac{\langle:\hat{n}_{\textbf{k}}\hat{n}_{\textbf{k}^\prime}:\rangle}{\langle\hat{n}_\textbf{k}\rangle \langle\hat{n}_{\textbf{k}^\prime} \rangle} = \frac{P_{\textbf{k},\textbf{k}^\prime}}{\langle\hat{n}_\textbf{k}\rangle \langle\hat{n}_{\textbf{k}^\prime} \rangle},
\end{equation}
where $\hat{n}_\textbf{k}$($\hat{n}_{\textbf{k}^\prime}$) is the momentum-space occupation of mode $\textbf{k}$($\textbf{k}^\prime$). Integrating the correlated atom counts over the selected momentum-space detection volumes associated with the four output ports $D_{\{1,2,3,4\}}$ yields the coincidence probabilities $P_{\textbf{p},\textbf{p}^\prime}$, $P_{\textbf{q},\textbf{q}^\prime}$, $P_{\textbf{p},\textbf{q}^\prime}$, and $P_{\textbf{q},\textbf{p}^\prime}$, providing direct measurement of the interferometric phase dependence.

\section{\label{Detection}Momentum-resolved detection and mode selection}
Following the detection and analysis procedures outlined in Ref.\,\cite{athreyaBell2026}, individual metastable $^4$He* atoms are detected using a microchannel plate (MCP) and delay-line detector (DLD) system located 848 mm below the trap. This system takes advantage of the approximately $\sim$19.8 eV internal energy of the metastable helium atoms, which is released upon detection. The detector has an estimated quantum efficiency of 20(2)\% \cite{kannanMeasurement2024}, with spatial and temporal resolutions of $120\,\mu\text{m}$ and $3\,\mu\text{s}$, respectively \cite{manningHanbury2010}. These resolutions correspond to momentum resolutions of approximately $\sim(4.5\times10^{-3})k_0$ in the transverse directions and $\sim(4.6\times10^{-4})k_0$ along the vertical direction. 

Atomic momenta are reconstructed from the measured arrival positions and times and subsequently transformed into the centre-of-momentum (COM) frame of each scattering halo. Owing to the relatively small number of scattered atoms per halo, the COM is instead determined from the corresponding parent condensates, which contain approximately $5 \times 10^3$ atoms and thus provide a more accurate estimate of the collision centre. This transformation is performed independently for each experimental realisation, thereby minimising broadening of the momentum distributions due to shot-to-shot fluctuations such as in-trap oscillations of the BEC. In addition, calibration runs are performed in which the initial double-halo state is generated, and the measured $g^{(2)}$ correlations across the halo are used to apply minor adjustments to the COM positions, yielding optimised estimates of the halo centres for each dataset. The datasets are further filtered based on the number of detected scattered atoms to ensure approximately constant mode occupancy and correlation amplitudes.

For the dual-resonant Bragg interferometer, detection windows are selected around the resonantly coupled momentum doublets in regions $A$ and $B$ of the scattering halo. The coupled momentum modes are centred approximately $\pm0.016\,\text{m/s}$ from the halo equator, with each detection window extending $0.005\,\text{m/s}$ around the corresponding mode centre. These detection ranges are chosen to maximise the detected pair signal while limiting averaging of the interferometric phase across the finite momentum acceptance of each output port. Across the selected detection windows, the phase variation arising from different particle trajectories is approximately 0.42 radians, resulting in only a small reduction of the measured interference contrast (see section\,\ref{visibility} for details).

The selected atoms satisfy the radial velocity constraint $0.9\leq v/v_r\leq1.1$, where $v_r\approx65\,\text{mm/s}$  is the velocity radius of the scattering halo. The four interferometer output ports $D_1,D_2,D_3,D_4$ correspond to the selected momentum-mode regions in halos $A$ and $B$. Joint detection probabilities are obtained by integrating correlated atom counts over the corresponding momentum-space detection volumes. Each scattering mode is defined by a cubic integration volume with side length approximately $0.025\times\text{k}_r$, where $\text{k}_r=mv_r/\hbar$, which provides a mode size sufficiently small to preserve momentum correlations while allowing several independent momentum modes to be sampled within each detection window.

\section{\label{g2s}Two-particle correlations of the double halo source}
\begin{figure}
    \centering
    \includegraphics[width=0.5\linewidth]{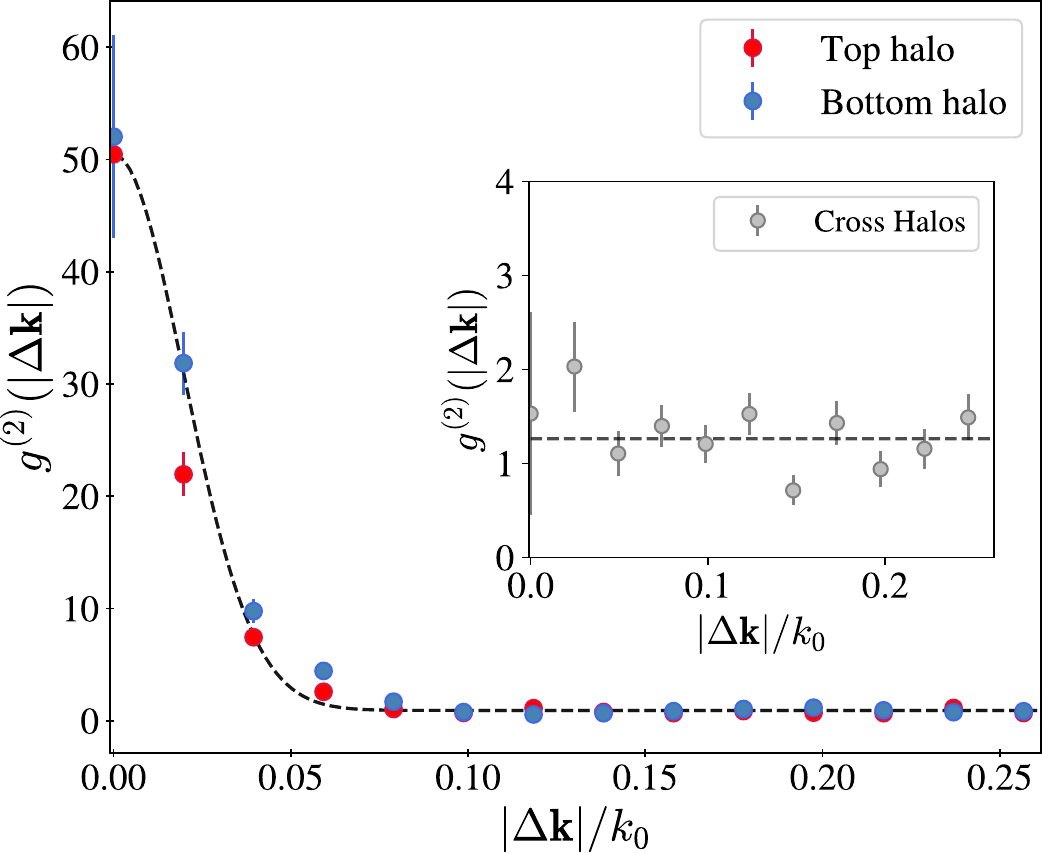}
    \caption{\textbf{Back-to-back correlations of the double halo source.} Measured second-order correlation functions $g^{(2)}$ for the initial double-halo state. The main panel shows the intra-halo correlations for the `Top'(red) and `Bottom'(blue) halos, with dashed curves showing Gaussian fits to the correlation peaks. The inset displays the cross-halo correlations, which remain close to the uncorrelated background, with a constant fit yielding $g^{(2)} \approx 1.3(2)$. Error bars denote shot-noise uncertainties.}
    \label{g2fig}
\end{figure}

The double-halo source is characterised through measurements of the second-order momentum correlation function \cite{hodgmanSolving2017},
\begin{equation}
    g^{(2)}(\Delta\textbf{k})\equiv g^{(2)}(\textbf{k},-\textbf{k}+\Delta\textbf{k}),
\end{equation}
evaluated within the momentum-space detection windows defined above (section\,\ref{Detection}). Figure \ref{g2fig} shows the measured back-to-back correlations for the initial double-halo state prior to the application of the dual-resonant Bragg interferometer. These measurements characterise the momentum-correlated atom-pair source produced by spontaneous four-wave mixing \cite{perrinObservation2007,perrinAtomic2008} and determine the mean mode occupancy relevant to the Bell-correlation measurements.

The measured intra-halo correlation functions exhibit pronounced peaks at $|\Delta\textbf{k}|=0$, with fit amplitudes of $g^{(2)}(0)=50.5(4)$ and $50.9(7)$ for the top(red) and bottom(blue) halos, respectively. Using the relation $g^{(2)}(0)\simeq 2+1/\bar{n}$\cite{hodgmanSolving2017}, the corresponding mean mode occupancy is estimated to be $\bar{n}\approx0.02$ for both halos, confirming operation well within the spontaneous, low-gain regime \cite{lewis-swanUltracold2016}.

The cross-halo correlation function, shown in the inset of Fig.\,\ref{g2fig}, remains close to the uncorrelated background level, $g^{(2)}\approx1$, demonstrating negligible correlations between atoms originating from different halos. Together, these measurements confirm that correlated atom pairs are generated predominantly within individual halos, while simultaneous independent pair production across both halos is strongly suppressed.

\section{\label{visibility}Visibility reduction of the Bell correlation function}
The Bell correlation function \cite{clauserProposed1969} is constructed from the measured joint detection probabilities as
\begin{equation}
    E(\phi_A,\phi_B) = \frac{P_{\textbf{p},\textbf{p}^\prime}+P_{\textbf{q},\textbf{q}^\prime}-P_{\textbf{p},\textbf{q}^\prime}-P_{\textbf{q},\textbf{p}^\prime}}{P_{\textbf{p},\textbf{p}^\prime}+P_{\textbf{q},\textbf{q}^\prime}+P_{\textbf{p},\textbf{q}^\prime}+P_{\textbf{q},\textbf{p}^\prime}} ,
    \label{Ecorr}
\end{equation}
which yields the expected phase dependence;
\begin{equation}
    E(\phi_A,\phi_B)=-V\cos(\phi_A+\phi_B),
    \label{Ecorr_theory}
\end{equation}
where $V$ is the Bell correlation interference visibility. For an ideal maximally entangled state and infinitesimal detection modes, the visibility is expected to approach unity. In the present experiment, the measured visibility is reduced by two dominant effects: higher-order Fock-state contributions to the generated atom-pair state and phase averaging arising from the finite momentum acceptance of the detection ports.

The reduction due to higher-order Fock-state (HOF) components is given by \cite{lewis-swanUltracold2016,lewis-swanProposal2015a},
\begin{equation}
    V_{\text{HOF}}=\frac{1+\bar{n}}{1+3\bar{n}}
\end{equation}
where $\bar{n}$ is the mean mode occupancy. Using the measured value $\bar{n}\approx 0.02$ obtained from the two-particle correlation measurements (section\,\ref{g2s}) yields
\begin{equation}
    V_{\text{HOF}}\approx0.962.
\end{equation}

A second contribution arises from integrating over the finite momentum extent of each detection port. Atoms within a given detection window acquire slightly different interferometric phases due to their distinct momentum trajectories, resulting in a small reduction of the observed fringe contrast. Following the treatment of \cite{kannanFramework2026}, this effect gives
\begin{equation}
    V_{\text{window}} = 0.992.
\end{equation}

Assuming these two effects are independent, the expected visibility is
\begin{equation}
V_{\text{expected}}=V_{\text{HOF}}V_{\text{window}}\approx0.962\times0.992\approx0.95
\end{equation}
This agrees well with the experimentally measured visibility of $V=0.93(2)$, indicating that the observed reduction from the ideal value of unity is quantitatively accounted for by higher-order pair production and finite detection-port integration.

\section{\label{error}Statistical Error Analysis}
The Bell correlation function is estimated from repeated experimental realizations of the dual-resonant interferometer. The low mode occupancy of the double-halo source results in the majority of experimental realizations containing at most a single detected correlated atom pair within the selected momentum detection windows. Consequently, each experimental realization contributes a binary outcome to the Bell correlation measurement, with the measured value of the correlation function $E$ (Eq.\eqref{Ecorr}) taking either $+1$ or $-1$. This binary outcome is a natural consequence of operating in the spontaneous low-gain regime and does not rely on any additional post-selection based on the detected atom number.
The statistical uncertainty is therefore evaluated using a binomial proportion estimator \cite{Cetinkaya-RundelOpenIntroStat2019}. The standard error of the measured proportion is calculated as $\text{SE} = \sqrt{\frac{p(1 - p)}{N}}$, where $p$ is the measured proportion of one of the two outcomes and $N$ is the sample size.

The same binary-outcome statistics apply to the measurements of the joint detection probabilities $P_{\textbf{k},\textbf{k}^\prime}$, where each realization corresponds to the detection or absence of a pair within a specified pair of output ports. Therefore, the uncertainties in both the Bell correlation function and the joint detection probabilities are evaluated using the corresponding binomial standard errors.

The visibility is extracted by fitting the measured Bell correlation function to the expected sinusoidal dependence, as in Eq.\eqref{Ecorr_theory}. The uncertainty in $V$ is determined from the covariance matrix of the least-squares fit, thereby incorporating the statistical uncertainties of the individual correlation measurements.

The CHSH-Bell parameter $S$ is evaluated from the four measured Bell correlation values \cite{clauserProposed1969,rarityExperimental1990},
\begin{equation}
    S=|E(\phi_A^{(1)},\phi_B^{(1)})-E(\phi_A^{(1)},\phi_B^{(2)})+E(\phi_A^{(2)},\phi_B^{(1)})+E(\phi_A^{(2)},\phi_B^{(2)})|.
\end{equation}
Assuming the four correlation measurements are statistically independent, the corresponding uncertainty is obtained by standard error propagation,
\begin{equation}
    \sigma_S=\sqrt{\sigma^2_{E_1}+\sigma^2_{E_2}+\sigma^2_{E_3}+\sigma^2_{E_4}}
\end{equation}
where $\sigma_{E_i}$ are the binomial uncertainties associated with the individual correlation measurements. The resulting uncertainty is reported for the experimentally measured CHSH-Bell parameter.

\end{document}